\documentclass[12pt]{article}
\usepackage{oldgerm}
\usepackage{yfonts}
\usepackage{euler}
\usepackage{graphics}
\usepackage{bm}
\usepackage{graphicx}
\usepackage{epstopdf}
\usepackage{amsmath}
\usepackage{amssymb}
\usepackage{amstext}
\usepackage{amscd}
\usepackage{amsfonts}
\usepackage{appendix}
\usepackage{color}
\usepackage{wasysym}
\usepackage{latexsym}
\DeclareMathAlphabet{\EuFrak}{U}{euf}{m}{n}
\DeclareMathAlphabet{\EuScript}{U}{eus}{m}{n}

\newcommand{\dQ}[1]{\frac{\partial {#1}}{\partial Q}}

\newcommand{\corch}[1]{\left[1+(q-1){#1}\right]}

\newcommand{\nd}{\noindent}

\newcommand{\be}{\begin{equation}}
\newcommand{\ee}{\end{equation}}
\newcommand{\ben}{\begin{eqnarray}}
\newcommand{\een}{\end{eqnarray}}

\title{{\bf q-Path entropy phenomenology for phase-space curves}}

\author{{D. J. Zamora$^1$ M. C. Rocca$^1$,
A. Plastino$^1$, and G. L. Ferri$^2$} \\
\small{$^1$ La Plata National University and
Argentina's National Research Council}\\
\small{(IFLP-CCT-CONICET)-C. C. 727, 1900 La Plata - Argentina}\\
$^2$ \small{Fac. de C. Exactas-National University La Pampa,} \\
\small{Peru y Uruguay, Santa Rosa, La Pampa, Argentina}}

\date{\today}

\begin{document}

\maketitle

\begin{abstract}
\noindent We describe the phenomenology of the classical q-path
entropy of a phase-space curve. This allows one to disclose an
entropic force-like   mechanism that is able to mimic some
phenomenological aspects of the strong force, such as confinement,
hard core, and asymptotic freedom.

\nd {\bf Keywords:} Phase-space curves; Entropic force; q-statistical mechanics

\end{abstract}

\newpage

\renewcommand{\theequation}{\arabic{section}.\arabic{equation}}

\setcounter{equation}{0}

\section{Introduction}

\setcounter{equation}{0}

\setcounter{equation}{0}
\subsection{q-Statistics}

The so called  q-statistical mechanics has been used in multiple applications in the last
 years \cite{tsallis,web,pre,epjb1,epjb2,epjb3,epjb4,epjb5,epjb6,epjb7,epjb8,epjb9}, being of  great
relevance for astrophysics, particularly for self-gravitating systems \cite{PP93,chava,lb,rosa}.
 Additionally, it was useful in variegated scientific fields. It
has originated several thousands of both papers and authors \cite{web}. Research on
its structural features is  important for astronomy, physics,
neurology, biology, economic sciences, etc.  Its success reaffirms
the idea that a great deal of physics derives from exclusively statistical considerations,
rather than mechanical ones. A foremost example lies in its application to high
energy physics. Here  q-statistics seems to adequately describe the transverse
momentum distributions of different hadrons \cite{tp11o,tp11,phenix}. In this sense, $q=1.15$ has acquired particular relevance \cite{tp11o,tp11,phenix}.

\subsection{Our goals}
\nd We describe in this work the phenomenology of classical q-path entropies for
arbitrary phase-space curves $\Gamma$, disclosing in such a way  interesting
features, like confinement and hard-cores.  Of course,  by
confinement we are reminded of the  phenomenon that impedes
isolation of color charged particles (quarks that cannot
be isolated singularly) and thus cannot be directly
detected, while by asymptotic freedom we make reference to  a property of some gauge
theories that generates particles' bonds to become
asymptotically weaker as distance diminishes, and at high momenta.
 Our curves-research will yield, surprisingly enough,  a
simple, classical  entropic  mechanism that is able to mimic some of  above cited  phenomena.
\vskip 3mm

\nd Remind that the
 entropic force is a phenomenological one that derives
from  statistical tendencies to entropic growth
\cite{pol1,pol2,verlinde,Wissner,dewar,curves,vis,to,hen}, without appealing  to  any
specific  underlying microscopic interaction.  The foremost
example is the elasticity of a freely-jointed polymer molecules
   \cite{pol1,pol2}.  It is remarkable that Verlinde has argued that gravity can also be understood
in terms of an entropic force \cite{verlinde}. \vskip 2mm

\nd Here we use q-path entropies  to demonstrate that
confinement can classically emerge from entropic forces, by appeal to
a quadratic Hamiltonian in phase-space. These Hamiltonians  are well know,
 classically and quantumly. For them, the
correspondence between classical and quantum mechanics is exact.
Unfortunately,  explicit formulas are not  trivial.\vskip 2mm \nd
Knowledge of quadratic Hamiltonians is of utility for investigating
more general Hamiltonians (and their associated
Schroedinger equations) in a semiclassical scenario. Quadratic
Hamiltonians are relevant  in partial differential equations:
 they   yield non trivial instances of wave propagation
phenomena. Quadratic Hamiltonians  also help in gaining insight into properties of more involved Hamiltonians
used in quantum theory.
\vskip 2mm

\nd We thus will appeal to quadratic Hamiltonians in a
classical environment so as  to learn whether some
interesting properties emerge concerning the entropic force
along phase-space curves, which will indeed be the fact. A similar previous analysis involving
the $q=1$, ordinary Boltzmann entropy has been reported in  \cite{curves}. $q \ne 1$ will be seen to add some interesting refinements.

The paper is organized as follows.
Section 2 reviews basic notions. In Section 3
we compute the q-path entropy and associated mean energy for the harmonic oscillator (HO). Section 4 deals with the concomitant energy-equipartition. Entropic forces linked to path entropies are the subject of
Section 5. We specialize this force in Section 6.
Conclusions are drawn in Section 7.

\section{Preliminaries}

\setcounter{equation}{0}

We consider a particle, moving in phase-space,  attached to a
harmonic oscillator-spring connected to the origin (the center of
HO-attraction).  We write the Hamiltonian in the fashion
\begin{equation}
\label{eq2.1}
H(P,Q)=P^2+Q^2,
\end{equation}
where both $P^2$ and $Q^2$ have the dimension of $H$.   Our environment is a Tsallis-one  \cite{tsallis} for which probability distributions are so-called q-exponentials, that  read  \cite{tsallis}

\be \label{qexponen}   e_q(x) = [1 +(1-q)x]^{1/(1-q)}. \ee

\nd The partition function in a Tsallis' scenario is defined as \cite{tsallis}
\be  Z(\beta)=\int_{-\infty}^{\infty} \int_{-\infty}^{\infty} \,e_q[-\beta H(P,Q)] dPdQ, \ee
or
\begin{equation}
\label{eq2.2}
Z(\beta)=\int_{-\infty}^{\infty} \int_{-\infty}^{\infty} {\corch{\beta H(P,Q)}}^{\frac{1}{1-q}} dPdQ,
\end{equation}
where $1\leq q<2$ is the non-extensibility parameter. Note that, if $q\rightarrow1$, the partition function reduces to the usual one , that means, the Gibbs-Boltzmann's partition function.
Of course, $\beta$ is the inverse temperature. In a Jaynes' MaxEnt information-theory context \cite{jaynes}, $\beta$ is a Lagrange multiplier that guarantees conservation of the mean energy, regarded as a  priori datum, when one maximizes the entropy (in order to obtain the appropriate probability distribution).  Jaynes' theory does not need neither ensembles nor heat baths \cite{jaynes}. \vskip 3mm

\nd Appealing  now to  the change of variables
\begin{equation}
\label{eq2.3}
U=P^2+Q^2, Q^{'}=\sqrt{U-P^2}
\end{equation}
we obtain
\begin{equation}
Z(\beta)=\pi \int_{0}^{\infty} {\corch{\beta U}}^{\frac{1}{1-q}} dU.
\label{ZU}
\end{equation}
Evaluating now (\ref{ZU}) we have (see ref.\cite{tr})
\begin{equation}
\label{eq2.5}
Z(\beta)=\frac{\pi}{\beta (q-1)} B\left[1,\frac{2-q}{q-1}\right],
\end{equation}
where $B[a,b]$ is the beta function. Thus,
\begin{equation}
\label{eq2.6}
Z(\beta)=\frac{\pi}{\beta (q-1)}
\frac{\Gamma(1)\Gamma\left(\frac{2-q}{q-1}\right)}
{\Gamma\left(\frac{1}{q-1}\right)},
\end{equation}
or equivalently,
\begin{equation}
Z(\beta)=\frac{\pi}{\beta (2-q).}
\label{Zinf}
\end{equation}
If $q\rightarrow1$, then
\begin{equation}
\label{eq2.8}
Z\rightarrow \frac{\pi}{\beta},
\end{equation}
and Eq. \ref{Zinf} thus reduces to the expression obtained
in \cite{curves} for the Gibbs-Boltzmann statistics. Similarly, for the mean value of the energy we have
\begin{equation}
\label{eq2.9}
<U>(\beta)=\frac{1}{Z}\int_{-\infty}^{\infty} \int_{-\infty}^{\infty} H(P,Q){\corch{\beta H(P,Q)}}^{\frac{1}{1-q}} dPdQ,
\end{equation}
or
\begin{equation}
<U>(\beta)=\frac{\pi}{Z} \int_{0}^{\infty} U{\corch{\beta U}}^{\frac{1}{1-q}} dU.
\label{UU}
\end{equation}
The result of (\ref{UU}) is
\begin{equation}
\label{eq2.11}
<U>(\beta)=\frac{\pi}{\beta^2 (q-1)^2 Z} B\left[2,\frac{3-2q}{q-1}\right],
\end{equation}
that can be recast as
\begin{equation}
\label{eq2.12}
<U>(\beta)=\frac{\pi}{\beta^2 Z} \frac{1}{(2-q)(3-2q)}.
\end{equation}
Replacing   above the value of $Z$ we obtain
\begin{equation}
<U>(\beta)=\frac{1}{\beta (3-2q)},
\label{Uinf}
\end{equation}
with the restriction $1\leq q<1.5$ in order to guarantee the non-divergence of $<U>$. When $q\rightarrow1$ we obtain
\begin{equation}
\label{eq2.14}
<U>\rightarrow \frac{1}{\beta},
\end{equation}
and for the entropy we have
\begin{equation}
S(\beta)=ln_{2-q}Z+Z^{q-1}\beta<U>,\label{s}
\end{equation}
where $ln_{q}(z)$ is the q-logarithm function  defined as
\begin{equation}
\label{eq2.16}
ln_q(z)=\frac{z^{1-q}-1}{1-q}.
\end{equation}
Appropriate replacing  in (\ref{s}) we are led to
\begin{equation}
\label{eq2.17}
S(\beta)=Z^{q-1}\left(\beta<U>+\frac{1}{q-1}\right)-\frac{1}{q-1},
\end{equation}
and
\begin{equation}
\label{eq2s}
S(\beta)=\left[\frac{\pi}{\beta(2-q)}\right]^{q-1}\frac{(2-q)}{(3-2q)(q-1)}-\frac{1}{q-1}.
\end{equation}
This result for  $S$ is valid for the interval $1<q<2$ and, of course, outside the poles, according
to Ref. \cite{tr}. Remember that the integrations here involved yield finite result only for  $1<q<2$, and otherwise diverge.

\setcounter{equation}{0}

\section{Path entropy}
We focus now on the concept of  path entropy \cite{curves}. The path is a phase-space curve $\Gamma$ parameterized by the variable $Q$. Following the procedure of \cite{curves}, we shall specialize equations \ref{ZU} and \ref{UU} to curves $\Gamma$. We first  define

\begin{equation}
Z(\beta,\Gamma)=\pi \int_{\Gamma}{\corch{\beta U(P,Q)}}^{\frac{1}{1-q}} dU(P,Q).
\label{zgamma}
\end{equation}
If we consider curves (parametrized by the independent variable
$Q$) passing through the origin, we have $P(0)=0$ and $Q=0$, and
consequently $U(0,0)=0$. Since the integrand in equation
\ref{zgamma} is an exact differential and the integral only depends
on the end point $Q_0$ we have
\begin{equation}
\label{eq3.2}
Z(\beta,Q_{0})=\pi \int_{0}^{U(Q_{0},P(Q_0))}
{\corch{\beta U(P,Q)}}^{\frac{1}{1-q}} dU(P,Q),
\end{equation}
or
\begin{equation}
Z(\beta,Q_{0})=\frac{\pi}{(2-q)\beta}
\{1-{\corch{\beta U(P(Q_{0}),Q_{0})}}^{\frac{2-q}{1-q}}\}
\label{zq}.
\end{equation}
If $Q_{0}\rightarrow\infty$ then $Z(\beta,Q_{0})$ reduces to
expression \ref{Zinf}.
Moreover if $q\rightarrow 1$, then
\begin{equation}
\label{eq3.4}
Z(\beta,Q_{0})\rightarrow\frac{\pi}{\beta}(1-e^{-\beta U(P(Q_{0}),Q_{0})}).
\end{equation}
In the same way, we have for the mean value of energy
\begin{equation}
\label{eq3.5}
<U>(\beta,\Gamma)=\frac{\pi}{Z(\beta,\Gamma)} \int_{\Gamma} U(P,Q){\corch{\beta U(P,Q)}}^{\frac{1}{1-q}} dU(P,Q),
\end{equation}
or equivalently,
\begin{equation}
\label{eq3.6}
<U>(\beta,Q_{0})=\frac{\pi}{Z(\beta,Q_{0})} \int_{0}^{U(Q_{0}P(Q_0))}
 U(P,Q){\corch{\beta U(P,Q)}}^{\frac{1}{1-q}} dU(P,Q)
\end{equation}
If we evaluate (\ref{eq3.6}) we obtain
\begin{multline}
\label{eq3.7}
<U>(\beta,Q_{0})=\frac{\pi}{Z(\beta,Q_{0})\beta^2} \left\{ \frac{1-{\corch{\beta U(P(Q_{0}),Q_{0})}}^{\frac{3-2q}{1-q}}}{(3-2q)(q-1)}- \right.\\
\left. \frac{1-{\corch{\beta U(P(Q_{0}),Q_{0})}}^{\frac{2-q}{1-q}}}{(2-q)(1-q)} \right\}
\end{multline}
and, simplifying the last expression,
\begin{multline}
<U>(\beta,Q_{0})=\frac{1}{\beta (q-1)} \left\{ -1+\frac{(2-q)}{(3-2q)} \frac{\{1-{\corch{\beta U(P(Q_{0}),Q_{0})}}^{\frac{3-2q}{1-q}}\}}{{\{1-{\corch{\beta U(P(Q_{0}),Q_{0})}}^{\frac{2-q}{1-q}}}\}}\right\}.
\label{uq}
\end{multline}
If $Q_{0}\rightarrow\infty$, then $<U>(\beta,Q_{0})$ reduces to expression (\ref{Uinf}).
If $q\rightarrow 1$, then
\begin{equation}
\label{eq3.9}
<U>(\beta,Q_{0})\rightarrow\frac{1-(1+\beta U)e^{-\beta U}}{(1-e^{-\beta U})\beta},
\end{equation}
which is  consistent with the results obtained in \cite{curves}.
Again, we can express  the entropy via the partition function and the mean value of energy, i.e.,
\begin{equation}
\label{eq3.10}
S(\beta,Q_{0})=ln_{2-q}Z(\beta,Q_{0})+Z(\beta,Q_{0})^{q-1}\beta<U>(\beta,Q_{0}),
\end{equation}
or, equivalently,
\begin{multline}
S(\beta,Q_0)=\frac{1}{q-1}\left\{\frac{(2-q)}{(3-2q)}\left[\frac{\pi}{(2-q)\beta}\right]^{(q-1)}\{1-{\corch{\beta U(P(Q_{0}),Q_{0})}}^{\frac{2-q}{1-q}}\}^{(q-2)}\right.\\
\left.\{1-{\corch{\beta U(P(Q_{0}),Q_{0})}}^{\frac{3-2q}{1-q}}\}-1\right\}.
\label{sq}
\end{multline}

Note here that, for  this $S-$expression, the same considerations
made above  for Eq.  (\ref{eq2s}) also hold. This is, the result for  $S$ is valid for the interval $1<q<2$ and, of course, outside the poles, according
to Ref. \cite{tr}. We reiterate
 that the integrations here involved yield finite result only for  $1<q<2$, and otherwise diverge.

\setcounter{equation}{0}

\section{Equipartition}
We find that
\begin{equation}
\label{eq4.1}
<Q^{2}>=\frac{1}{Z}\int_{-\infty}^{\infty} \int_{-\infty}^{\infty} Q^{2}{\corch{\beta (P^2+Q^2)}}^{\frac{1}{1-q}} dPdQ,
\end{equation}
or
\begin{equation}
\label{eq4.2}
<Q^{2}>=\frac{\pi}{2Z} \int_{0}^{\infty} U{\corch{\beta U}}^{\frac{1}{1-q}} dU,
\end{equation}
while along the curve $\Gamma$,
\begin{equation}
\label{eq4.3}
<Q^{2}>(\beta,\Gamma)=\frac{\pi}{2Z(\beta,\Gamma)} \int_{\Gamma} U{\corch{\beta U}}^{\frac{1}{1-q}} dU.
\end{equation}
Thus,
\begin{equation}
\label{eq4.4}
<Q^2>(\beta,Q_{0})=\frac{\pi}{2Z(\beta,Q_{0})} \int_{0}^{U(Q_{0},P(Q_0))}
U{\corch{\beta U}}^{\frac{1}{1-q}} dU.
\end{equation}
This integral yields
\begin{multline}
\label{eq4.5}
<Q^2>(\beta,Q_{0})=\frac{\pi}{2Z(\beta,Q_{0})\beta^2} \left\{ \frac{1-{\corch{\beta U(P(Q_{0}),Q_{0})}}^{\frac{3-2q}{1-q}}}{(3-2q)(q-1)}- \right.\\
\left. \frac{1-{\corch{\beta U(P(Q_{0}),Q_{0})}}^{\frac{2-q}{1-q}}}{(2-q)(1-q)} \right\},
\end{multline}
and then
\begin{equation}
\label{eq4.6}
<Q^2>(\beta,Q_{0})=\frac{<U>(\beta,Q_{0})}{2},
\end{equation}
that is,
\begin{equation}
\label{eq4.7}
<Q^2>(\beta,Q_{0})=<P^2>(\beta,Q_{0})=\frac{<U>(\beta,Q_{0})}{2},
\end{equation}
which, for $Q_{0}\rightarrow\infty$ gives
\begin{equation}
\label{eq4.8}
<Q^2>=<P^2>=\frac{<U>}{2}=\frac{1}{2(3-2q)\beta},
\end{equation}
the q-equipartition recipe \cite{ppt}. In this phenomenological environment, 
it obviously sets an upper bound for $q$, namely, $q <3/2$, that applies only to the q-path entropy
 and not at all to the wide world of q-non additivity.
\setcounter{equation}{0}

\section{Entropic force}

According to \cite{verlinde,Wissner}, the entropic force $F_e$ is given by
\begin{equation}
\label{eq6.1}
F_{e}=\frac{1}{\beta} \frac{\partial S}{\partial Q_0}.
\end{equation}
Since the end-point of the curve  $Q_0$ is arbitrary, we may call it $Q$.  In our case this is
\begin{equation}
\label{eq6.2}
F_e=\frac{Z^{q-2}}{\beta} \dQ{Z}+(q-1)Z^{q-2} \dQ{Z}<U>+Z^{q-1}\dQ{<U>}.
\end{equation}\vskip 3mm

\nd See that $Z$ and $<U>$ were already  determined in \ref{zq} and \ref{uq}, respectively.
Note that these quantities are functions of $\beta$ and $Q$.
Recall that when we write $U$ we refer to $U(P(Q),Q)$.
Accordingly, we have
\begin{equation}
\label{eq6.3}
\dQ{Z}=\pi \dQ{U} \corch{\beta U}^{\frac{1}{1-q}}=2\pi
Q \corch{\beta U}^{\frac{1}{1-q}},
\end{equation}
and
\begin{multline}
\label{eq6.4}
\dQ{<U>}=\frac{(2-q)}{(3-2q)(q-1)}.\frac{2Q}{\left\{1-{\corch{\beta U}}^\frac{2-q}{1-q}\right\}} \left\{ (3-2q) \corch{\beta U}^\frac{2-q}{1-q}\right.\\
\left.-(2-q)\corch{\beta U}^\frac{1}{1-q} \frac{\{1-{\corch{\beta U}}^{\frac{3-2q}{1-q}}\}}{{\{1-{\corch{\beta U}}^{\frac{2-q}{1-q}}}\}}\right\}.
\end{multline}
Rearranging terms and properly  adding or subtracting when necessary, one we finds

\begin{multline}
\label{eq6.5}
\dQ{<U>}=\frac{2Q(2-q)}{(q-1)}.\frac{\corch{\beta U}^\frac{1}{1-q}}{\left\{1-{\corch{\beta U}}^\frac{2-q}{1-q}\right\}}\\
\left\{\corch{\beta U}-1+1-\frac{(2-q)}{(3-2q)}\frac{\{1-{\corch{\beta U}}^{\frac{3-2q}{1-q}}\}}{{\{1-{\corch{\beta U}}^{\frac{2-q}{1-q}}}\}}\right\},
\end{multline}
while (\ref{eq6.5}) can be written as

\begin{multline}
\label{eq6.6}
\dQ{<U>}=2Q(2-q)\frac{\corch{\beta U}^\frac{1}{1-q}}{\left\{1-{\corch{\beta U}}^\frac{2-q}{1-q}\right\}}
\{\beta U-\beta<U>\}.
\end{multline}
Thus, we have for $F_e$ the expression
\begin{multline}
\label{eq6.7}
F_e=\left(\frac{\pi}{\beta}\right)^{q-1}2Q(2-q)^{2-q}\frac{\corch{\beta U}^{\frac{1}{1-q}}}{\left(1-{\corch{\beta U}}^{\frac{2-q}{1-q}}\right)^{2-q}}
\left\{\beta U+\frac{1}{q-1}\right.\\
\left.-\frac{1}{(q-1)}\frac{(2-q)^2}{(3-2q)} \frac{\{1-{\corch{\beta U}}^{\frac{3-2q}{1-q}}\}}{{\{1-{\corch{\beta U}}^{\frac{2-q}{1-q}}}\}}\right\}.
\end{multline}
This is of the form

\be   F_e= K(q, \beta, U[Q,P]) Q,  \label{restora}\ee
and will be either attractive or repulsive. The entropic force is here ''harmonic''-like  too. It is not conservative because it depends on $P$ and not only on position. In the examples we have considered in this paper, K is positive. Where does it come from? We might argue as follows: we want our particle to traverse a given path in phase space. Such path may connect points corresponding to different HO-energies. As a response to our traveling particle, a non conservative force arises that opposes its motion, in \`a la Chatelier fashion.
\vskip 3mm

\nd If $q\rightarrow 1$,
\begin{equation}
\label{eq6.8}
\dQ{Z}\rightarrow2\pi Qe^{-\beta U},
\end{equation}

\begin{equation}
\label{eq6.9}
\dQ{<U>}\rightarrow \frac{2Qe^{-\beta U}}{1-e^{-\beta U}}\left[\frac{\beta U}{1-e^{-\beta U}}-1\right],
\end{equation}
and
\begin{equation}
\label{eq6.10}
F_{e}\rightarrow2Q\beta U  \frac{e^{-\beta U}}{(1-e^{-\beta U})^2}.
\end{equation}
Clearly, $K$ is positive and the force repulsive.
For $q$ close to unity, a first order approximation yields

\begin{equation}
\label{eq6.11}
Z\approx\pi U,
\end{equation}

\begin{equation}
\label{eq6.12}
\dQ{Z}\approx2\pi Q (1-\beta U)\approx2\pi Q,
\end{equation}

\begin{equation}
\label{eq6.13}
<U>\approx0,
\end{equation}

\begin{equation}
\label{eq6.14}
\dQ{<U>}\approx2Q(1-\beta U)\approx2Q,
\end{equation}
and
\begin{equation}
\label{eq6.15}
F_{e}\approx\frac{\pi}{\beta}(\pi U)^{q-2}2Q+(\pi U)^{q-1}2Q\approx(\pi U)^{q-1}2Q(\frac{1}{\beta U}+1).
\end{equation}
Finally, the entropic force simplifies to
\begin{equation}
\label{eq6.16}
F_{e}\approx(\pi U)^{q-1}\frac{2Q}{\beta U},
\end{equation}
with $K$ positive and, for   $q\rightarrow1$,
\begin{equation}
\label{eq6.17}
F_{e}\approx\frac{2Q}{\beta U},
\end{equation}
in agreement with  \cite{curves}. We have carefully explored the whole ($q,\,p$) plane and discovered that effects of $F_e$ are noticeable only in a small region of it.

\vskip 3mm \nd We can still replace $U$ in the equations above using $U=P^2+Q^2$. One has
\begin{multline}
\label{eq7.1}
F_e=\left(\frac{\pi}{\beta}\right)^{q-1}
2Q(2-q)^{2-q}\frac{\corch{\beta(P^2+Q^2)}^{\frac{1}{1-q}}}
{\left(1-{\corch{\beta(P^2+Q^2)}}^{\frac{2-q}{1-q}}\right)^{2-q}}\\
\left\{\beta(P^2+Q^2)+\frac{1}{q-1}\right.\\
\left.-\frac{1}{(q-1)}\frac{(2-q)^2}{(3-2q)}
\frac{\{1-{\corch{\beta(P^2+Q^2)}}^{\frac{3-2q}{1-q}}\}}
{{\{1-{\corch{\beta(P^2+Q^2)}}^{\frac{2-q}{1-q}}}\}}\right\}.
\end{multline}

\begin{figure}[h!]
\begin{center}
\includegraphics[width=8.6cm]
{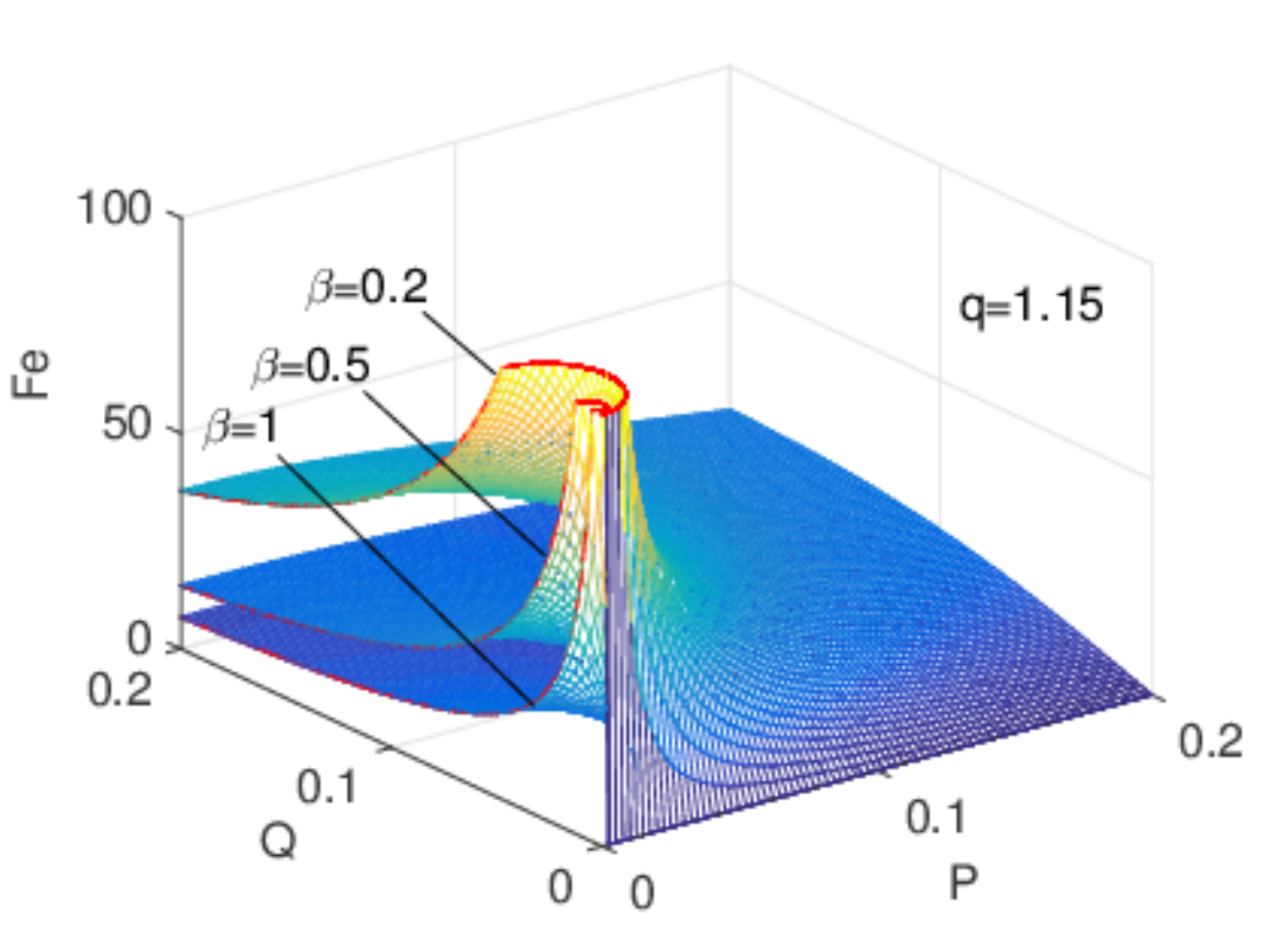} \caption{Behavior of the entropic force
with $T$. We note that the entropic force grows with the
temperature,
 as one should expect.} \label{figura1}
\end{center}
\end{figure}

\begin{figure}[h!]
\begin{center}
\includegraphics[width=8.6cm]
{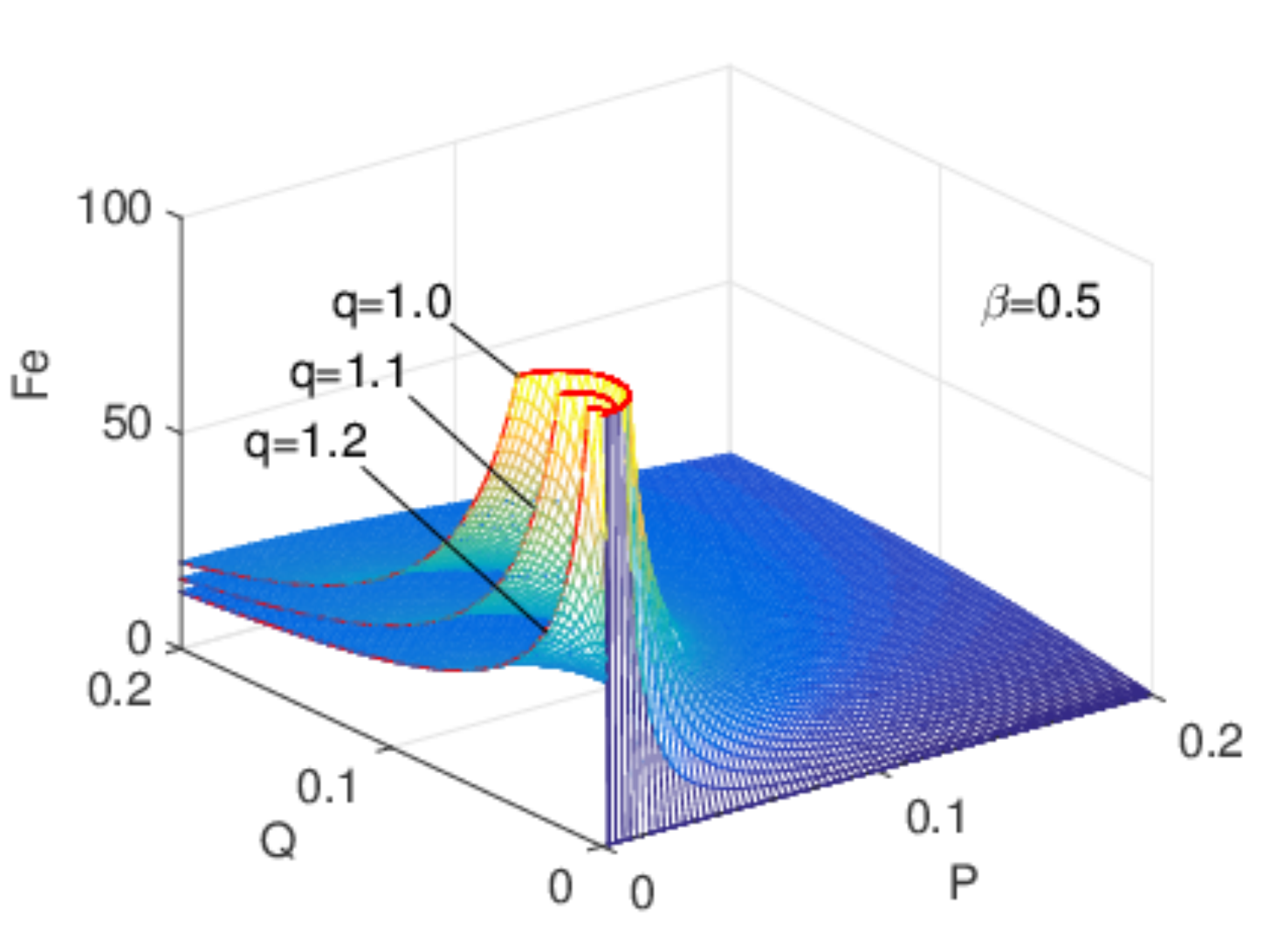} \caption{Behavior of the entropic force with $q$.
We see that the entropic force diminishes as $q$ grows, which constitutes a new result.} \label{figura2}
\end{center}
\end{figure}

\nd We understand by force's confinement "that there is just a  region where
the force is significant". We see that the entropic force is in this sense confined to just a small region of phase space. This confinement effect i) grows with $q$  and ii) leads to asymptotic freedom (zero force) outside such a region.
We note also that is the particle lies inside the
potential-barrier, it will be hard  for it  to escape (no tunneling in classical physics). The peak of the entropic force becomes more pronounced as $q$ increases. We point out that,
as the kinetic energy, associated to the momentum, grows, so does confinement. These observations are illustrated by Figs. 1 and 2.

\newpage

\setcounter{equation}{0}

\section{The effects of the HO well}
Obviously, our particle also feels an harmonic  force. Let us add its effect to that of $F_e$. We have
\begin{equation}
\label{eq8.1}
F_{HO}=-\frac{1}{2}\dQ{<Q^2>}=-\frac{1}{4}\dQ{<U>}.
\end{equation}
In our case this becomes

\begin{equation}
\label{eq8.2}
F_{HO}=-\frac{Q}{2}(2-q)\frac{\corch{\beta U}^\frac{1}{1-q}}
{\left\{1-{\corch{\beta U}}^\frac{2-q}{1-q}\right\}}
\{\beta U-\beta<U>\}.
\end{equation}
Note that, for $q\rightarrow1$,
\begin{equation}
\label{eq8.3}
F_{HO}\rightarrow-\frac{Q}{2}\frac{e^{-\beta U}}{(1-e^{-\beta U})}\left(\frac{\beta U}{1-e^{-\beta U}}-1\right),
\end{equation}
or
\begin{equation}
\label{eq8.4}
F_{HO}\rightarrow\frac{Q}{2}\frac{e^{-\beta U}}{(1-e^{-\beta U})^2}(1-\beta U-e^{-\beta U}).
\end{equation}
The total force that the particle feels is
\begin{equation}
\label{eq8.5}
F_T=F_e+F_{HO}=Z^{q-2} \dQ{Z}\left(\frac{1}{\beta}+(q-1)<U>\right)+\dQ{<U>}\left(Z^{q-1}-\frac{1}{4}\right),
\end{equation}
and then:
\begin{multline}
\label{eq8.6}
F_T=\left(\frac{\pi}{\beta}\right)^{q-1}
2Q(2-q)^{2-q}\frac{\corch{\beta(P^2+Q^2)}^{\frac{1}{1-q}}}
{\left(1-{\corch{\beta(P^2+Q^2)}}^{\frac{2-q}{1-q}}\right)^{2-q}}\\
\left\{\beta(P^2+Q^2)+\frac{1}{q-1}\right.\\
\left.-\frac{1}{(q-1)}\frac{(2-q)^2}{(3-2q)}
\frac{\{1-{\corch{\beta(P^2+Q^2)}}^{\frac{3-2q}{1-q}}\}}
{{\{1-{\corch{\beta(P^2+Q^2)}}^{\frac{2-q}{1-q}}}\}}\right\}-\\
\frac{Q}{2}(2-q)\frac{\corch{\beta(P^2+Q^2)}^\frac{1}{1-q}}
{\left\{1-{\corch{\beta(P^2+Q^2)}}^\frac{2-q}{1-q}\right\}}\\
\left[\beta(P^2+Q^2)+\frac{1}{(q-1)}
\left(1-\frac{(2-q)}{(3-2q)}
 \frac{\{1-{\corch{\beta(P^2+Q^2)}}^{\frac{3-2q}{1-q}}\}}
 {{\{1-{\corch{\beta(P^2+Q^2)}}^{\frac{2-q}{1-q}}}\}}\right)\right]
\end{multline}

In order to appreciate the meaning of the above equation, consider its limit for $q\rightarrow1$,
\begin{equation}
\label{eq8.7}
F_T\rightarrow\frac{Q}{2}\frac{e^{-\beta U}}{(1-e^{-\beta U})^2}(1+3\beta U-e^{-\beta U}).
\end{equation}
The absolute value of the force grows linearly with the distance $Q$. It also grows with $T$.

\begin{figure}[h!]
\begin{center}
\includegraphics[width=8.6cm]
{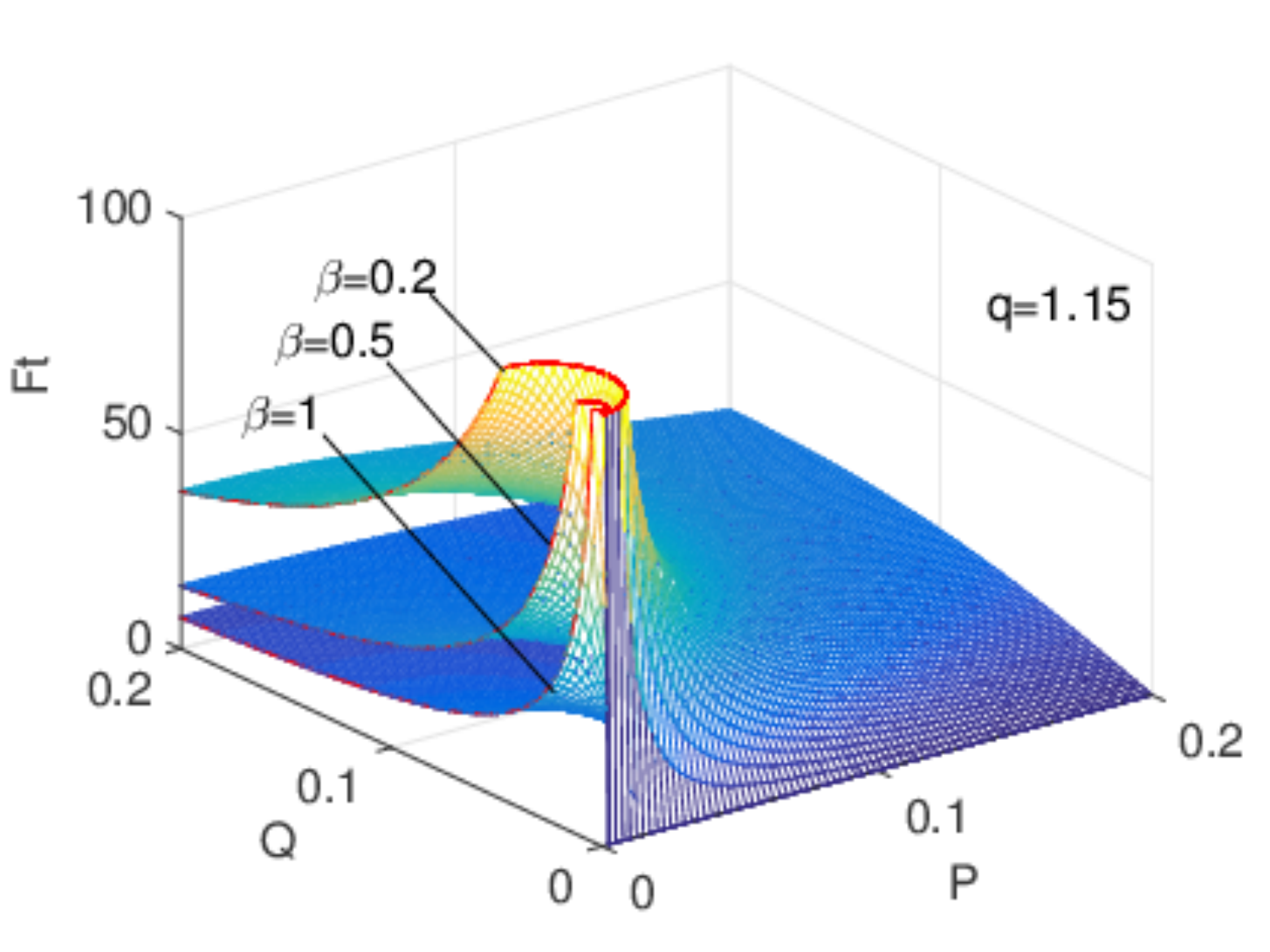} \caption{Total force associated to the path entropy as a function of $T$. It grows as the temperature increases, as one should expect.} \label{figura3}
\end{center}
\end{figure}

\begin{figure}[h!]
\begin{center}
\includegraphics[width=8.6cm]
{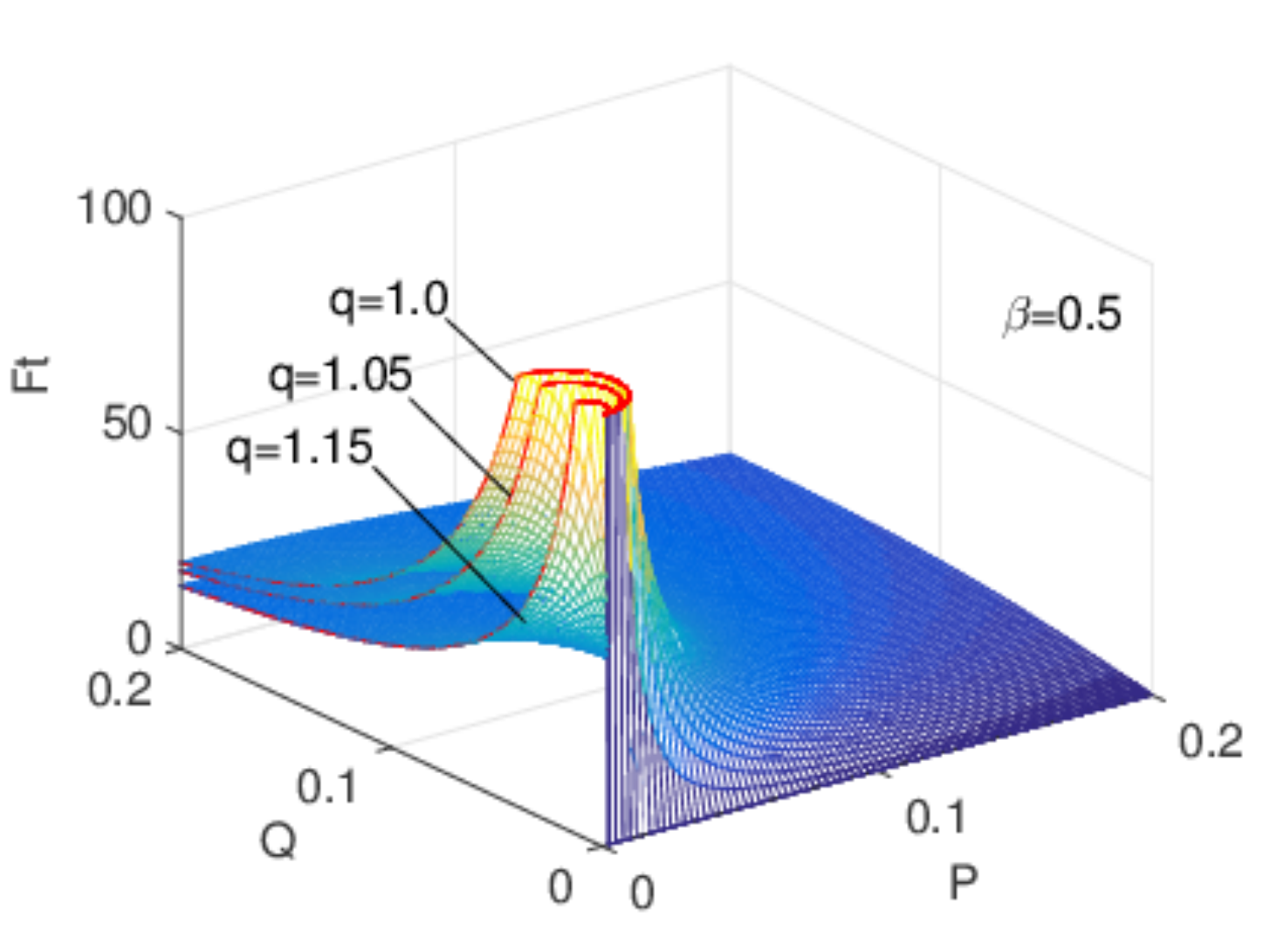} \caption{Total force associated to the path entropy as a function of $q$.
 It diminishes as $q$ grows.} \label{figura4}
\end{center}
\end{figure}

\nd Notice that, as above, the total force is confined (different from zero) to just a small region of phase space. Such confinement effect i) grows with $q$  and ii) leads to asymptotic freedom (zero total force) outside such a region. Once again, the peak of the total force becomes more pronounced as $q$ increases. Also, as $F_{total}$ grows, so does confinement. We see that, here, the entropic force is a manifestation of the entropic change along a curve in phase space. As our particle traverses this path, it feels the action of $F_e$. These considerations are illustrated by Figs. 3 and 4.

\newpage

\setcounter{equation}{0}

\section{Conclusions}

In this paper we have introduced the notion of classical  q-path entropy and described its behavior for the classical harmonic oscillator (HO). In other words, we have described the phenomenology of the q-path entropy for the
classical HO. One main facet of this phenomenology is the existence of
 q-entropic force effects. We discovered that one can
classically mimic,  with the q-entropic force, aspects of another phenomenology:
 that of   strong forces.\vskip 2mm

\begin{itemize}

\item Although we
discussed arbitrary phase space curves $\Gamma$ , in general our effects do not depended upon
the specific $\Gamma$ chosen. Our  q-phenomenology appears  reasonable  because a q-equipartition theorem holds.
\vskip 2mm

\item We
 discovered that  the q-entropic force diverges at specific small areas of phase space (hard-core effect), vanishing outside those areas (confinement plus asymptotic  freedom). The size of these areas is q-dependent.
The ''hardness'' of the core also depends upon q. Adding a harmonic well does not modify this
phenomenology.\vskip 2mm

\item  Also,  the entropic force grows with $q$, an interesting result. The more non-additive the environment becomes, the stronger the entropic force.   This force also increases with temperature, which  constitutes a reasonable result.

\item The q-dependence that we  have just described justifies the
appeal to a non-additive environment that is the leite motif of
this effort.

\end{itemize}

\newpage

\renewcommand{\thesection}{\Alph{section}}

\renewcommand{\theequation}{\Alph{section}.\arabic{equation}}

\setcounter{section}{1}


\newpage

\begin{thebibliography}{99}

\bibitem{tsallis}  M. Gell-Mann and C. Tsallis, Eds. {\it Nonextensive Entropy:
Interdisciplinary applications}, Oxford University Press, Oxford,
2004.

\bibitem{web} See http://tsallis.cat.cbpf.br/biblio.htm for a
regularly updated bibliography on the subject.

\bibitem{pre} P-Jizba, J. Korbel,  V. Zatloukal,
Phys. Rev. E {\bf 95}, 022103 (2017);
L. S. F. Olavo,
Phys. Rev. E  {\bf 64}, 036125 (2001).


\bibitem{epjb1} I. S. Oliveira:
Eur. Phys. J. B {\bf 14}, 43 (2000)
\bibitem{epjb2} E. K. Lenzi , R. S. Mendes:
Eur. Phys. J. B {\bf 21}, 401 (2001)
\bibitem{epjb3} C. Tsallis:
Eur. Phys. J. A {\bf 40}, 257 (2009)
\bibitem{epjb4} P. H. Chavanis:
Eur. Phys. J. B {\bf 53}, 487 (2006)
\bibitem{epjb5} G. Ruiz , C. Tsallis:
Eur. Phys. J. B {\bf 67}, 577 (2009)
\bibitem{epjb6} P. H. Chavanis , A. Campa:
Eur. Phys. J. B {\bf 76}, 581 (2010)
\bibitem{epjb7} N. Kalogeropoulos:
Eur. Phys. J. B {\bf 87}, 56 (2014)
\bibitem{epjb8} N. Kalogeropoulos:
Eur. Phys. J. B {\bf 87}, 138 (2014)
\bibitem{epjb9} A. Kononovicius, J. Ruseckas:
Eur. Phys. J. B {\bf 87}, 169 (2014)

\bibitem{PP93} A. R. Plastino, A. Plastino,  Phys.  Lett. A {\bf 174}, 384 (1993)

\bibitem{jaynes} E.T. Jaynes, Phys. Rev. {\bf 106} (1957) 620; {\bf 118} (1961) 171; Papers on
probability, statistics and statistical physics, edited by R. D. Rosenkrantz,
Reidel, Dordrecht, Holland, 1983; L. Brillouin, Science and Information
Theory, Academic Press, New York (1956); WT Grandy, Jr., and PW
Milonni, Physics and probability: Essays in honor of E.T. Jaynes, Cambridge University Press, Cambridge, England, 1993.


\bibitem{ppt}  A. R. Plastino, A. Plastino, C. Tsallis,  Journal of Physics A  {\bf 27}  (1994) 5707.


\bibitem{chava} P. H. Chavanis, C. Sire, Physica A {\bf 356} (2005)
419; P.-H. Chavanis, J. Sommeria, Mon. Not. R. Astron. Soc. {\bf
296}, 569 (1998).


\bibitem{lb} D. Lynden-Bell,  R. M. Lynden-Bell,  Mon. Not. R. Astron. Soc. {\bf
181}, 405 (1977).



\bibitem{tp11o} C. Tsallis, {\it Introduction to Nonextensive Statistical Mechanics},
Springer, Berlin, 2009.

\bibitem{tp11}   F. Barile et al. (ALICE Collaboration),
EPJ Web Conferences {\bf 60}, 13012 (2013); B. Abelev et al.
(ALICE Collaboration), Phys. Rev. Lett. {\bf 111}, 222301 (2013);
Yu. V.Kharlov (ALICE Collaboration), Physics of Atomic Nuclei {\bf
76}, 1497 (2013); ALICE Collaboration,  Phys. Rev. C 91, (2015)
 ATLAS Collaboration, New J. Physics {\bf 13}, 024609 (2011);
053033; CMS Collaboration, J. High Energy Phys. {\bf 05},064 (2011);
 CMS Collaboration, Eur. Phys. J. C {\bf 74}, 2847 (2014)

\bibitem{phenix}  A. Adare et al (PHENIX Collaboration), Phys. Rev.
D {\bf   83},  052004 (2011);  PHENIX Collaboration,  Phys. Rev. C
{\bf   83},  024909 (2011); PHENIX Collaboration, Phys. Rev. C
{\bf   83}, 064903 (2011); PHENIX Collaboration,  Phys. Rev. C
{\bf   84}, 044902 (2011)

\bibitem{pol1} H.W. de Haan, G.W. Slater, Phys. Rev. E {\bf 87},  042604  (2013)


\bibitem{pol2} M.F. Maghrebi, Y. Kantor, M. Kardar, Phys. Rev. E {\bf 86},  061801  (2012)

\bibitem{verlinde} E. Verlinde: J. High Energy Physics.  {\bf 04},  29 (2011)



\bibitem{Wissner}  A.D. Wissner-Gross, C.E. Freer, Phys. Rev. Lett.  {\bf 110},  168702  (2013)


\bibitem{dewar}   R.C. Dewar, Entropy  {\bf 11}, 931 (2009);
R.C. Dewar, A. Maritan, arXiv:1107.1088.

\bibitem{rosa} A. C. P. Rosa Jr. et al. Physica A {\bf 392}, 6079-6083 (2013)

\bibitem{vis} M. Visser. JHEP {\bf 10} 140, (2011).

\bibitem{to} T. Wang. Phys. Rev. D {\bf 81} 104045, (2010).

\bibitem{hen} S. H. Hendi and A. Sheykhi. Int. J. Theor. Phys
{\bf 51} 1125, (2012).

\bibitem{curves} M. Rocca, A. Plastino, G.L. Ferri: Physica A {\bf 393}, 244 (2014)


\bibitem{tr} I. S. Gradshteyn and I. M. Rizhik: ''Table of Integrals
Series and Products''. Academic Press (1965).


\end{thebibliography}
\end{document}